\documentclass[fleqn,twoside]{article}
\usepackage{espcrc2}

\usepackage{graphicx}
\usepackage[figuresright]{rotating}

\newcommand{\AmS}{{\protect\the\textfont2
  A\kern-.1667em\lower.5ex\hbox{M}\kern-.125emS}}

\title{3D Position Sensitive XeTPC for Dark Matter Search}

\author{
J.~Angle\address[uf]{Department of Physics, University of Florida, Gainesville, FL 32611, USA},
E.~Aprile\address[cu]{Department of Physics, Columbia University, New York, NY 10027, USA}, 
F.~Arneodo\address[lngs]{Gran Sasso National Laboratory, Assergi, L'Aquila, 67010, Italy}, 
L.~Baudis\address[rwth]{Department of Physics, RWTH Aachen University, Aachen, 52074, Germany}, 
A.~Bernstein\address[llnl]{Lawrence Livermore National Laboratory, 7000 East Ave., Livermore, CA 94550, USA}, 
A.~Bolozdynya\address[case]{Department of Physics, Case Western Reserve University, Cleveland, OH 44106, USA}, 
L.~Coelho\address[coimbra]{Department of Physics, University of Coimbra, R. Larga, 3004-516, Coimbra, Portugal}, 
E.~Dahl\addressmark[case], 
L.~DeViveiros\address[brown]{Department of Physics, Brown University, Providence, RI 02912, USA},  
A.~Ferella\addressmark[lngs], 
L.~Fernandes\addressmark[coimbra], 
S.~Fiorucci\addressmark[brown], 
R.J.~Gaitskell\addressmark[brown], 
K.-L.~Giboni\addressmark[cu], 
R.~Gomez\address[rice]{Department of Physics, Rice University, Houston, TX, 77251, USA}, 
R.~Hasty\address[yale]{Department of Physics, Yale University, New Haven, CT 06511, USA}, 
J.~Kwong\addressmark[case], 
J.A.M.~Lopes\addressmark[coimbra], 
N.~Madden\addressmark[llnl], 
A.~Manalaysay\addressmark[uf], 
A.~Manzur\addressmark[yale], 
D.~McKinsey\addressmark[yale], 
M.E.~Monzani\addressmark[cu], 
K.~Ni\addressmark[yale] \thanks{Presented at the 7th UCLA Symposium on \textit{Sources and Detection of Dark Matter and Dark Energy in the Universe}. (E-mail: kaixuan.ni@yale.edu)},
U.~Oberlack\addressmark[rice], 
J.~Orboeck\addressmark[rwth], 
G.~Plante\addressmark[cu],  
J.~Santos\addressmark[coimbra], 
P.~Shagin\addressmark[rice], 
T.~Shutt\addressmark[case], 
P.~Sorensen\addressmark[brown], 
C.~Winant\addressmark[llnl], 
M.~Yamashita\addressmark[cu]
        }

\begin{document}

\begin{abstract}
The technique to realize 3D position sensitivity in a two-phase xenon time projection chamber (XeTPC) for dark matter search is described. Results from a prototype detector (XENON3) are presented.
\vspace{1pc}
\end{abstract}

\maketitle

\section{Introduction}
\label{}
The XENON dark matter search experiment~\cite{xenon01} uses time projection chamber (TPC) to search for the hypothetical WIMP dark matter particles. The detector (see the most recent status at \cite{yamashita:ucla06}) consists a bulk of liquid xenon (LXe) as the target for WIMP interactions. The target is also self-shielded from the external background events, mostly gamma rays. The external gamma rays have more chance to interact near the edge and surface of the LXe target. A fiducial volume cut of the edge and surface events will dramatically reduce the background rate and improve the detector's sensitivity for dark matter search. A fiducial volume cut will also help to remove events from the regions (edge or surface) with irregular or non-uniform electric fields. The signals of events from these regions can be similar to the WIMP signals and make them difficult to be rejected without 3D position sensitivity. A position sensitive detector will also have the capability to distinguish neutron interactions by their multiple scatters in the target, while a WIMP is very unlikely to make more than one scatter due to its tiny interaction cross-section. 

\section{3D Position Localization}
\label{}

The XENON experiment uses a two-phase (liquid/gas) xenon detector. An event in the detector produces two signals: a prompt direct scintillation light (S1) and a delayed proportional scintillation light (S2). The delayed S2 signal is from the ionization electrons that are drifted from the event site in the liquid to the gas phase. The event $Z$ position is calculated from the electron drift velocity (about 2~$\rm\mu s/mm$ at 1 kV/cm drift field) times the drift time.

\begin{figure}[htbp]
\centering
\includegraphics[height=5cm]{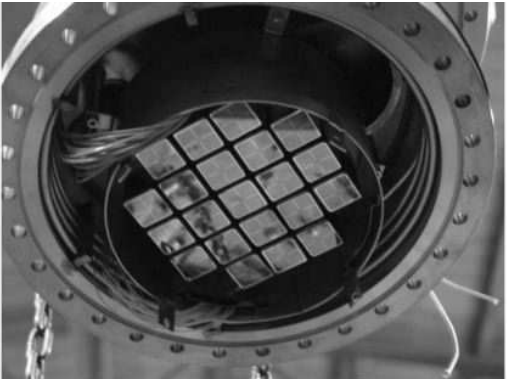} 
\caption{A 21-PMT array installed in the XENON3 detector.}
\label{xenon3}
\end{figure}

The diffusion of the drifted electrons in LXe is very small, which gives a very localized $X\&Y$ positions for the S2 signal. Using an array of photon-detectors on top of the S2 signal, the $X\&Y$ positions can be reconstructed from the S2 signal distributions in these photon-detectors. During the R\&D phase of XENON experiment, a detector (called XENON3) was constructed with 21 Hamamatsu R9288 PMTs (1-inch-square each) installed in the gas phase, right on top of the structure where proportional scintillation occurs. Fig.~\ref{xenon3} shows the structure of the 21-PMT array. The array covers the 5-cm-radius surface area of the LXe target.

To reconstruct the $X\&Y$ positions, a simulated S2 map is produced for comparing with the actual S2 distributions from the detector. The simulated S2 distributions on the 21 PMTs are produced for each $\rm1\times1~mm^2$ pixel in the 5-cm-radius sensitive area, where proportional scintillation light occurs. A minimum-$\chi^2$ method is used for the comparing. The minimum-$\chi^2$ method calculates the $\chi^2$ value as in equation \ref{eq:mchi}. $S_i$ and $s_i$ are the measured and expected (from simulation) S2 signals (in number of photoelectrons) for the $i$th PMT. $M$ is the total number of PMTs in the top PMT array (for XENON3 detector, $M$ = 21). 

\begin{equation}
  \chi^2(x,y) = \sum_{i=1}^{M} \frac{\left[S_i - s_i(x,y)\right]^2}{\sigma_i^2}
  \label{eq:mchi}
\end{equation}

Here $\sigma_i^2$ is the uncertainties from both the measured signal $S_i$ and simulated expectation signal $s_i$. If the simulation has enough statistics, the major contribution to $\sigma_i^2$ is the measured signal fluctuation, which includes the statistical fluctuations of photoelectron emission ($\sigma_{pe,i}$) from the PMT's photocathode and its gain fluctuation ($\sigma_{g_i}$). It can be approximately written as in equation \ref{eq:sigma_xy}. $\sigma_{pe,i}$ is simply equal to $\sqrt{S_i}$, providing $S_i$ is sufficiently large. $\sigma_{g_i}$ was measured for each PMT based on its single photoelectron spectrum, where $g_i$ is the gain of that PMT. 

\begin{equation}
\label{eq:sigma_xy}
\sigma^2_i = \sigma^2_{pe,i}\left[1 + (\sigma_{g_i}/g_i)^2\right]=S_i\left[1 + (\sigma_{g_i}/g_i)^2\right]
\end{equation}

The $\chi^2$ value is computed for all possible $X\&Y$ positions, in 1~mm$^2$ pixels. The minimum value, $\chi^2_{min}$, corresponds to the reconstructed event $X\&Y$ positions.

\section{Results}
\label{}
The $Z$ position, calculated from the electron drift time, is quite precise. The resolution is better than 1 mm. Due to the solid-angle effect of light collection and the electron-negative impurities in liquid xenon, both S1 and S2 signals are dependent on $Z$-position. The precise $Z$ positions thus can be used to correct the signals. The sub-mm resolution also allows a very precise fiducial volume cut based on $Z$.

\begin{center}
\begin{figure}[htbp]
\includegraphics[height=6cm]{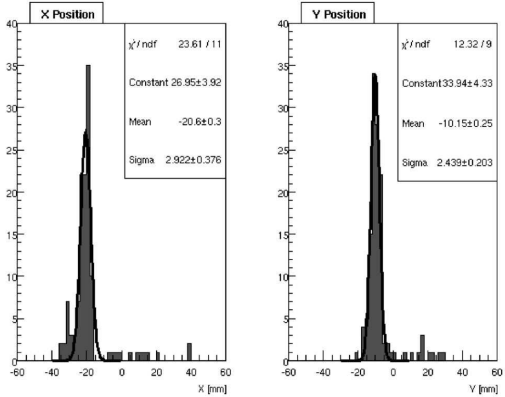} 
\caption{Recontructed $X\&Y$ positions of a $\alpha$ source located at (X = -24 mm, Y = -10 mm) in the XENON3 detector. The reconstructed $X$ position has a small offset. The position resolutions for both $X$ and $Y$ are less than 3 mm ($\sigma$).}
\label{alpha}
\end{figure}
\end{center}

With a localized 5.3~MeV $\alpha$ source in the detector, the $X\&Y$ position resolutions were measured to be around 3 mm ($\sigma$) (see Fig. \ref{alpha}). The $X\&Y$ position resolutions are not as good as that for $Z$. But they are sufficient for making fiducial volume cut in $X\&Y$. 

\begin{center}
\begin{figure}[htbp]
\includegraphics[height=6cm]{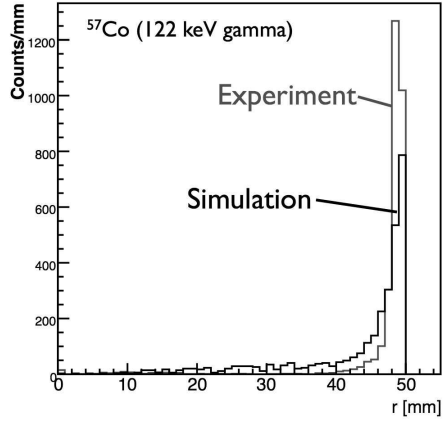} 
\caption{Radial position distribution for 122 keV $\gamma$-ray interacting in the XENON3 detector. Reconstructed positions from experiment are compared with simulations.}
\label{Co57}
\end{figure}
\end{center}

To verify the position sensitivity for events near the edge, an external low energy $\gamma$-ray source ($\rm^{57}Co$) was used. $\rm^{57}Co$ emits mainly 122 keV $\gamma$ rays. With its small attenuation length in liquid xenon, most of the 122 keV $\gamma$ rays interact near the detector's edge. Fig. \ref{Co57} shows the reconstructed  radial position distribution of the 122 keV $\gamma$-ray events, compared with expected distribution from a Geant4 simulation. The distributions from experiment and simulation match quite well, while the reconstructed positions have a tendency to the edge. This is very possibly due to the less coverage of PMTs near the edge.

The two-phase xenon detectors have a good gamma background rejection efficiency, based on the ratio of S2 and S1 signals, as demonstrated in \cite{Aprile:PRL06}. The S2/S1 value for nuclear recoils (from WIMPs) are much smaller than that from electron recoils (background gammas). The S2 signal is proportional to the number of electrons extracted from the liquid to the gas. In a realistic detector, the non-uniformity of the electric field near the edge usually gives an insufficient electron extraction (``edge effect"), resulted a smaller S2/S1 value, which makes the electron recoils leaking into the nuclear recoil region, as reported in \cite{yamashita:xesat05}.

\begin{figure}[htbp]
\centering
\includegraphics[height=5.8cm]{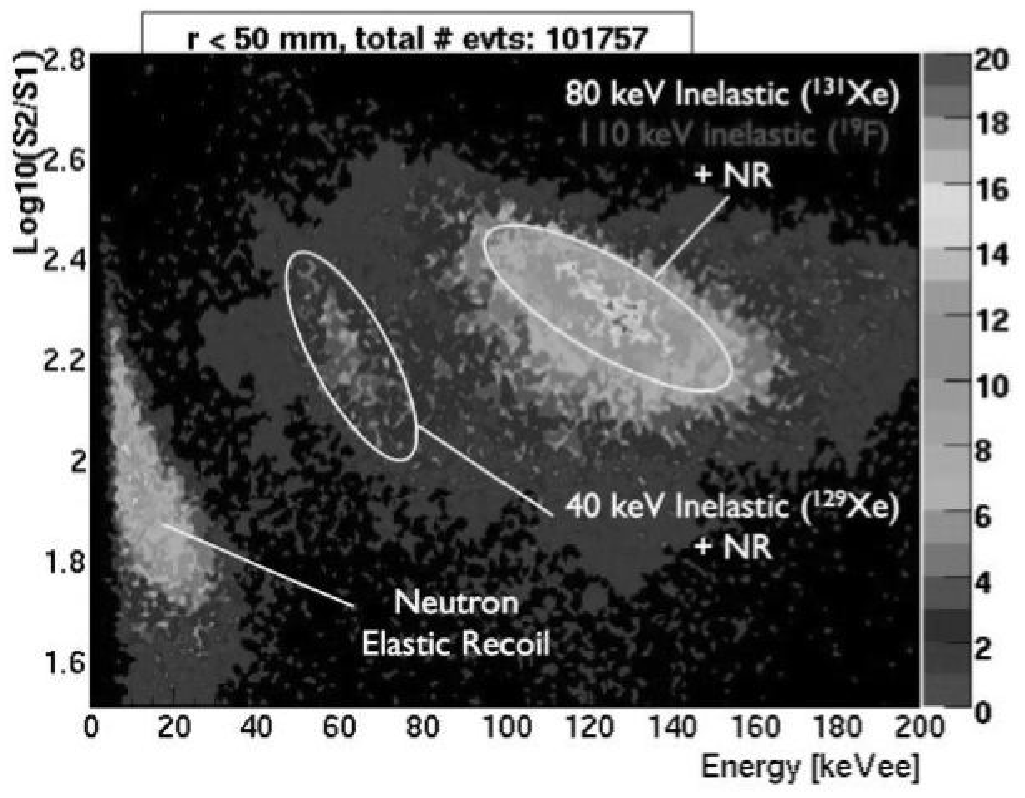} \\
\includegraphics[height=5.8cm]{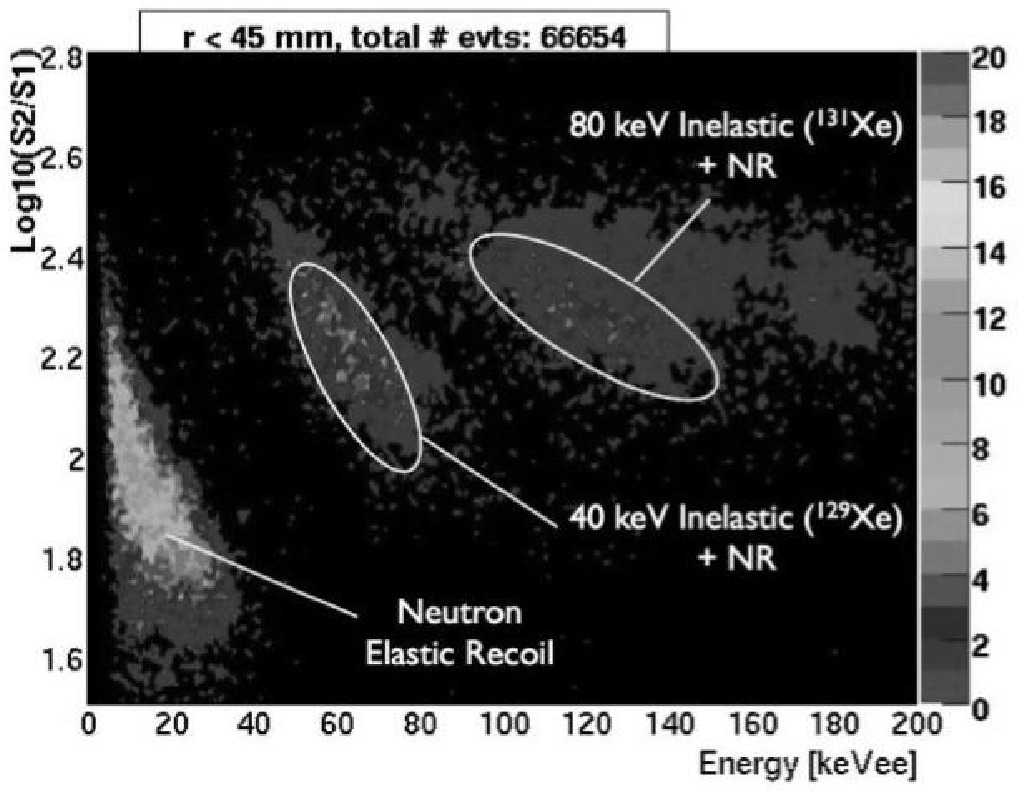} 
\caption{S2/S1 as a function of energy for events from neutron interactions in the XENON3 detector, without radial position cut (top) and with 5 mm radial position cut from edge (bottom). The energy scale (keVee: keV electron equivalent) is calibrated with 122 keV gamma rays. Note: color version of these two plots can be found in the talk at: http://www.physics.ucla.edu/hep/dm06/talks/ni.pdf)}
\label{s2s1_n}
\end{figure}

The XENON3 detector was exposed to a neutron source and the ``edge effect" was also observed, as in Fig. \ref{s2s1_n} (top). The neutron elastic recoil events are clearly separated from the electron recoils. Neutrons also interact inelastically and produce meta-stable states from $^{129}$Xe and $^{131}$Xe in the liquid xenon target. The $^{129m}$Xe and $^{131m}$Xe emit 40 keV and 80 keV gammas. Neutrons also make inelastic scattering on $^{19}$F in the PTFE material, surrounding the liquid xenon target. The PTFE is used to improve the scintillation light collection. The $^{19m}$F produces 110 keV gammas, mostly interacting near the edge. It's clear from Fig.~\ref{s2s1_n} (top) that a portion of these edge events have smaller S2/S1 value, due to the ``edge effect". A fiducial volume cut out of 5 mm from the edge significantly reduces the number of edge events and makes a better electron/nuclear recoil discrimination, as shown in Fig. \ref{s2s1_n} (bottom).

In a larger scale detector (e.g. 100 kg of LXe), the capability of rejecting neutron background becomes more necessary. Neutrons, unlike $\gamma$-ray background, will make elastic scattering on the target and produce nuclear recoils, the same as those from WIMPs. It's not possible to reject those neutron events based on S2/S1 ratio. But neutrons have a much larger interaction cross-section than the WIMPs and they can make multiple scatters in the target. A 3D Position sensitive detector, such as the one described in the current work, will be able to reject those background neutron events with multiple scatters.


A larger scale detector (XENON10) has been constructed and deployed underground for dark matter search \cite{yamashita:ucla06}. The same technique discussed here for the 3D position realization will be applied to XENON10. 

\section{Acknowledgments}
This work was funded by NSF Grants No. PHY-03-02646 and No. PHY-04-00596, and DOE Grant No. DE-FG02-91ER40688.



\end{document}